\documentclass[letter]{aa}  

\usepackage{graphicx}
\usepackage{subcaption}
\usepackage{txfonts}
\usepackage{hyperref}
\usepackage{xcolor}
\usepackage{ulem}
\usepackage{tikz,xcolor}
\usepackage{csquotes}
\usepackage{siunitx}
\usepackage{longtable}
\usepackage[para]{threeparttable}
\usepackage[para]{threeparttablex}
\usepackage{placeins}
\usepackage{multicol}
\DeclareSIUnit{\wavelength}{$\lambda$}
\DeclareSIUnit{\jansky}{Jy}
\DeclareSIUnit{\parsec}{pc}
\usepackage{scalerel}
\usepackage{tikz}
\usepackage{upgreek}
\usetikzlibrary{svg.path}
\usepackage{booktabs}
\definecolor{orcidlogocol}{HTML}{A6CE39}
\tikzset{
  orcidlogo/.pic={
    \fill[orcidlogocol] svg{M256,128c0,70.7-57.3,128-128,128C57.3,256,0,198.7,0,128C0,57.3,57.3,0,128,0C198.7,0,256,57.3,256,128z};
    \fill[white] svg{M86.3,186.2H70.9V79.1h15.4v48.4V186.2z}
                 svg{M108.9,79.1h41.6c39.6,0,57,28.3,57,53.6c0,27.5-21.5,53.6-56.8,53.6h-41.8V79.1z M124.3,172.4h24.5c34.9,0,42.9-26.5,42.9-39.7c0-21.5-13.7-39.7-43.7-39.7h-23.7V172.4z}electronically
                 svg{M88.7,56.8c0,5.5-4.5,10.1-10.1,10.1c-5.6,0-10.1-4.6-10.1-10.1c0-5.6,4.5-10.1,10.1-10.1C84.2,46.7,88.7,51.3,88.7,56.8z};
  }
}

\newcommand\orcidicon[1]{\href{https://orcid.org/#1}{\mbox{\scalerel*{
\begin{tikzpicture}[yscale=-1,transform shape]
\pic{orcidlogo};
\end{tikzpicture}
}{|}}}}

\begin{document}

   \title{Variability of Sagittarius\,A* at 3\,GHz on minute-scale with MeerKAT}

\author{
    K.\,Kaur\inst{\ref{mpifr}, \ref{aifa}} \and
    I.\,Rammala-Zitha\inst{\ref{mpifr}} \and
    A.\,Basu\inst{\ref{dza},\ref{tls},\ref{mpifr}} \and
    G.\,Witzel\inst{\ref{mpifr}} \and
    M.\,Wielgus\inst{\ref{iaa}} \and
 V. Balakrishnan\inst{\ref{cfa},\ref{mpifr}} \and
 E.D. Barr\inst{\ref{mpifr}} \and
 A. Brunthaler\inst{\ref{mpifr}} \and
 S. Buchner\inst{\ref{sarao}} \and
 D.J. Champion\inst{\ref{mpifr}} \and
 M.\,Hoeft\inst{\ref{tls}} \and
 S. Khan\inst{\ref{mpifr}} \and
 H.-R.\,Kl\"ockner\inst{\ref{mpifr}} \and
 C.\,K\"onig\inst{\ref{mpifr}} \and
 M. Kramer\inst{\ref{mpifr}} \and
 V. Venkatraman Krishnan\inst{\ref{mpifr}} \and
 Y.K.\, Ma\inst{\ref{mpifr}} \and
 S.A.\, Mao\inst{\ref{mpifr}} \and
 P.V. Padmanabh\inst{\ref{aei},\ref{luh}} \and
 S.\,Ranchod\inst{\ref{mpifr}} \and
 S.S.\,Sridhar\inst{\ref{skao}} \and
 J.D.\,Wagenveld\inst{\ref{mpifr}} \and
 R.S.\,Wharton\inst{\ref{mpifr}} \and
 O.\,Wucknitz\inst{\ref{mpifr}}
}

\institute{
    Max-Planck-Institut f\"ur Radioastronomie, Auf dem H\"ugel 69, 53121 Bonn, Germany \label{mpifr}
    \and
    Argelander Institute for Astronomy, University of Bonn, Germany \label{aifa}
    \and
    Deutsches Zentrum f\"ur Astrophysik, Postplatz 1, 02826 G\"orlitz, Germany \label{dza}
    \and
    Th\"uringer Landessternwarte, Sternwarte 5, 07778 Tautenburg, Germany \label{tls}
    \and
    Instituto de Astrofísica de Andalucía-CSIC, Glorieta de la Astronomía s/n, E-18008 Granada, Spain \label{iaa}
    \and
   Center for Astrophysics | Harvard \& Smithsonian, Cambridge, MA 02138-1516, USA \label{cfa}
    \and
    South African Radio Astronomy Observatory. 2 Fir Street, Black River Park, Observatory 7925, South Africa \label{sarao}
   \and
    Max-Planck-Institut f\"ur Gravitationsphysik (Albert Einstein Institute), 30167 Hannover, Germany \label{aei}
   \and
   Leibniz Universität Hannover, 30167 Hannover, Germany \label{luh}
   \and
    SKA Observatory, Jodrell Bank, Lower Withington, Macclesfield, SK11 9FT, United Kingdom \label{skao}
}

   \date{}

\abstract{
\textit{Context.} The supermassive black hole Sagittarius\,A* (Sgr\,A*) exhibits temporal and spectral variability across the electromagnetic spectrum. However, variability at radio frequencies 
below $\sim 5$\,GHz for timescales shorter than a day remains largely unexplored.

\textit{Aims.} We investigate the variability of Sgr\,A* at $2.79$\,\textrm{GHz} on short timescales (1\,min), to probe an under-explored regime of its emission process.

\textit{Methods.} 
Through point-source model fitting in the \textit{uv}-domain, we analyse the flux density variation of Sgr\,A* over an 8\,h observation.

\textit{Results.} We detect flux variation on a few tens of minute timescale with a modulation index of 6.11\%, a mean flux density of ($827 \pm 0.1_{\mathrm{stat}} \pm 33_{\mathrm{sys}}) \, \mathrm{mJy}$, and a mean spectral slope of $0.08\pm0.03$. Furthermore, we measure the slope of the structure function of the observed light curve as $0.81 \pm 0.05$ with a characteristic timescale of about 120\,min. 

\textit{Conclusions.} Our study at low radio frequencies is a critical step toward constraining the physical mechanisms that drive Sgr~A*'s variable emission and its spectral energy distribution.
Our study suggests that variability at centimetre and millimetre wavelengths is likely more closely related than previously thought. 
}


\authorrunning{K.\,Kaur et al.}

\maketitle

\nolinenumbers
\section{Introduction}
The compact source Sagittarius\,A* (Sgr\,A*) at the centre of our Galaxy is an ideal candidate for studying supermassive black holes (SMBHs) due to its proximity to Earth. A detailed understanding of Sgr\,A* would provide insight into the interactions of SMBHs at centres of distant low-luminosity galaxies with their environments.
\citet{brownandlo} observed its variability at radio frequencies (2695 and 8085 MHz) on both long (years) and short (one day to months) time scales. Multi-wavelength variability has also been observed, with frequent flaring 
in X-ray \citep[e.g.,][]{baganoff}, near-infrared (NIR) \cite[e.g.,][]{genzel2003, ghez2004}, and radio bands \citep[e.g.,][]{falcke1999, 2022ApJ...930L..19W}. The details of the physical mechanisms driving the variability, and the connection between the variability at different wavelengths remain unclear. 

At radio wavelengths, variability is believed to consist of two modes \citep{2006ApJ...650..189Y, 2021MNRAS.505.3616M}: -- (1) a flaring component (typically observed on timescales of hours), possibly linked to jets, outflows, magnetic reconnection events, or episodic accretion, and (2) a more stable quiescent component that persists across all timescales, likely associated with a radiatively inefficient accretion disk \citep[see][]{2003ANS...324..435Q,2003ApJ...598..301Y}. The rapidly varying flaring component superimposes on the quiescent component, and therefore the radio measurements capture the combined flux from both states \citep{2021MNRAS.505.3616M}. Moreover, the spectral energy distribution (SED) shows an increase in flux density from radio to sub-millimetre waveband with a positive spectral slope peaking near $10^4$\,GHz, the so-called ``sub-millimetre bump'' \citep[see e.g.,][]{zylka1995, falcke1998}, and dropping off sharply in the near-infrared regime. \cite{falcke1999} reported variability at 2.3 and 8.3 GHz on timescales of 50 to 200 days, while \cite{2004AJ....127.3399H} identified epochs of varying flux and spectral index, suggesting the presence of distinct emission states. Variability studies covering a broad frequency range, between 19 and 100\,GHz, indicate that the timescale increases at lower frequencies, possibly originating from a mildly relativistic outflow \citep{2015A&A...576A..41B}.

We present the first analysis of short-term variability in Sgr\,A* using the MeerKAT radio telescope \citep{2009IEEEP..97.1522J} at 2.79\,GHz. This relatively unexplored low-frequency regime offers a valuable window into the emission processes. These observations mark the first light curve obtained from our ongoing 200-hrs observing campaign dedicated to Sgr\,A* as part of the MPIfR-MeerKAT Galactic Plane Survey (MMGPS; \citealt{mmgps}).

With an intrinsic size of $\sim$11\,mas at 2.79\,GHz (using equation~3 in \citealt{refId0}), the light crossing time is expected to be $\sim$14 h. We find in our 8\,h data, minute-to-hour scale variability. We show that the variability at these low frequencies is similar to those found in millimetre observations with the Atacama Large Millimetre/sub-millimetre Array (ALMA) \citep{2022ApJ...930L..19W}. Our study also demonstrates the power of MeerKAT in studying the variability of Sgr\,A* with high fidelity, and paves the way for conducting such investigations in further detail.

\section{Temporal properties of Sgr\,A* at 2.79\,GHz}
\label{sec:results_discussion}

\subsection{Light Curve and Spectral Index}

In this Letter, we have analysed light curves of Sgr\,A* between 2.65 and 3\,GHz using data obtained with MeerKAT on 21 March 2024 between UTC 02:17 and 10:07, and investigate flux densities obtained from the \textit{uv}-domain. Summary of the observing strategy, data calibration, and flux density extraction are described in Appendix~\ref{sec:obs}. A detailed data processing is described in a separate paper (Rammala-Zitha et al., in preparation).

\begin{figure}
    \centering
    \includegraphics[width=\linewidth]{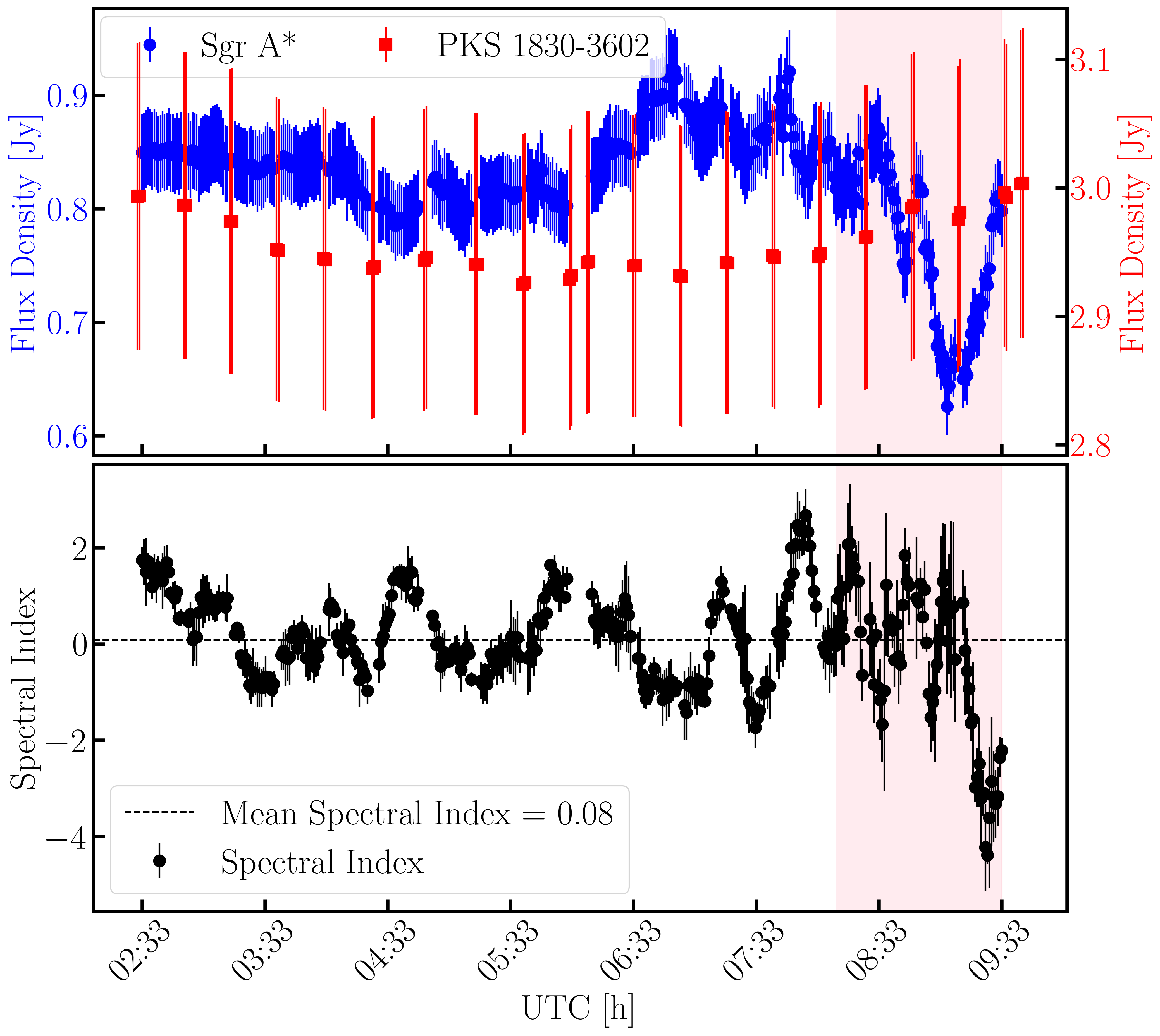} 
    \caption{\textbf{Top panel}: Total intensity light curve of Sgr\,A* (top panel) in bins of 1\,min (blue dots) and the phase calibrator PKS 1830-3602 (red squares) on 21 March 2024 at 2.79~GHz. The error bars include systematic and $1\sigma$ statistical uncertainties. \textbf{Bottom panel}: Spectral index ($\alpha$) variation of Sgr A* over time, derived from sub-band flux measurements. Error bars reflect statistical uncertainties. The dashed line represents the mean $\alpha = 0.08$ (full). The pink shaded region indicates the light curve after 08:13\,UTC  
    showing rapid decrease followed by increase in flux density.} 
    \label{fig:light_curve}
\end{figure}

Fig.~\ref{fig:light_curve} (top panel) shows the total intensity light curve binned into 1\,min intervals over the available bandwidth, while sub-band light curves are shown in Fig.~\ref{fig:light_curve_sub-band}. The sub-band light curves follow a similar trend to the total intensity light curve, with the highest frequency (2.87\,GHz) leading the lowest frequency (2.70\,GHz). This is further quantified in Section \ref{sec:time_lag}. The shaded interval in all light curve plots correspond to the time samples after $t >$\,08:13\,UTC, where both the flux density and spectral index vary rapidly; we discuss this interval further in Section \ref{sec:discussion}. From here on, we refer to the light curve spanning the full observation as ``full” and the light curve excluding samples in the shaded region as ``excluded”. 

We measure the mean flux density of $(827 \pm 0.1_{\mathrm{stat}} \pm 33_{\mathrm{sys}}) \, \mathrm{mJy}$ (full), and $(842 \pm 0.2_{\mathrm{stat}} \pm 34_{\mathrm{sys}}) \, \mathrm{mJy}$ (excluded).
These measurements are consistent with the flux density of $\rm 818\pm11\,mJy$ measured from the multi-frequency synthesis image, and with literature values  \citep[e.g][at 3\,GHz]{2015ApJ...802...69B}.
The modulation index, the ratio of standard deviation to mean flux density, is found to be $6.11 \pm 0.38\%$ (full) and $3.50 \pm 0.14\%$ (excluded), where the uncertainty is estimated as the standard deviation of bootstrap-resampled modulation index values. 

To assess the gain stability of the calibration, we plot the light curve of the gain calibrator PKS\,1830$-$3602 (the red squares) in Fig.~\ref{fig:light_curve} (top panel), in 1\,min bin. In contrast to Sgr\,A*, the calibrator light curve is flat, with a modulation index of 0.8\% of the mean flux. This is consistent with a non-variable gain calibrator, confirming calibration stability, and ruling out gain variations as plausible cause for the observed variability for Sgr\,A*. Further statistical tests to validate the robustness of the light curve are presented in Appendix~\ref{appendix:tests}.

We also use the four sub-bands images ($\sim$57\,MHz wide) to derive the spectral index. At each timestamp, the flux densities in the four sub-bands were fitted to obtain the spectral index ($S_\nu \propto \nu^\alpha$, where $S_\nu$ is the flux density at frequency $\nu$).
Fig.~\ref{fig:light_curve} (bottom panel) shows the time series of spectral index ($\alpha$) in 1\,min bins, ranging from $-4$ to $+2$ (full) and $-2$ to $+2$ (excluded).  
The mean is \(\alpha = 0.08 \pm 0.03\) (full) and \(\alpha = 0.20 \pm 0.02\) (excluded), consistent with literature measurements \citep[e.g][]{Iwata,2001ARA&A..39..309M,2004AJ....127.3399H}. Fig.~\ref{fig:spectral_index_vs_flux} shows the spectral index as a function of flux density, exhibiting a bimodal pattern: below 0.75\,Jy (coinciding with a dip in the temporal feature towards the end of the observation) the spectrum is steep, while above this threshold no distinctive trend is observed.

\subsection{Time Lag between sub-bands} \label{sec:time_lag}

We measure the mean spectral index close to zero, hinting at the possibility of emission near the transition between optically thick and thin synchrotron emission. In such a scenario, variability can appear at different times across frequencies, making it important to quantify possible delays \citep{1966Natur.211.1131V}. To investigate this, we performed a cross-correlation analysis between the light curves 
at 2.70 and 2.87\,GHz and find a time lag of $826\pm97$\,s (see Appendix~\ref{appendix B}).

In the optically thick regime, variability could be driven by adiabatic expansion.
Assuming expansion of a uniform, spherical blob of synchrotron-emitting plasma of relativistic electrons with energy index $p=2$, defined as $n(E) \propto E^{-p}$\citep{2021MNRAS.505.3616M}, the expected time lag between the two frequencies (using equation\,2 in \citealt{Iwata}) is $\approx 460$\,s, significantly lower than what we derive (see Fig.~\ref{fig:time_lag_parabola}). For electrons with strong radiative losses, where $p = 4$, the time lag is expected to be even lower ($\approx390$\,s). Considering a lower $p = 1.5$, that could arise from reconnection heating, the time lag is expected to be
$\approx490$\,s. Thus, our measurement differs by up to a factor of 2, and therefore, variability at 2.79\,GHz is unlikely to be caused by adiabatic expansion.

\subsection{Structure Function} \label{sec:sf_acf}

\begin{figure}[htbp]
    \centering
    \includegraphics[width=\linewidth]{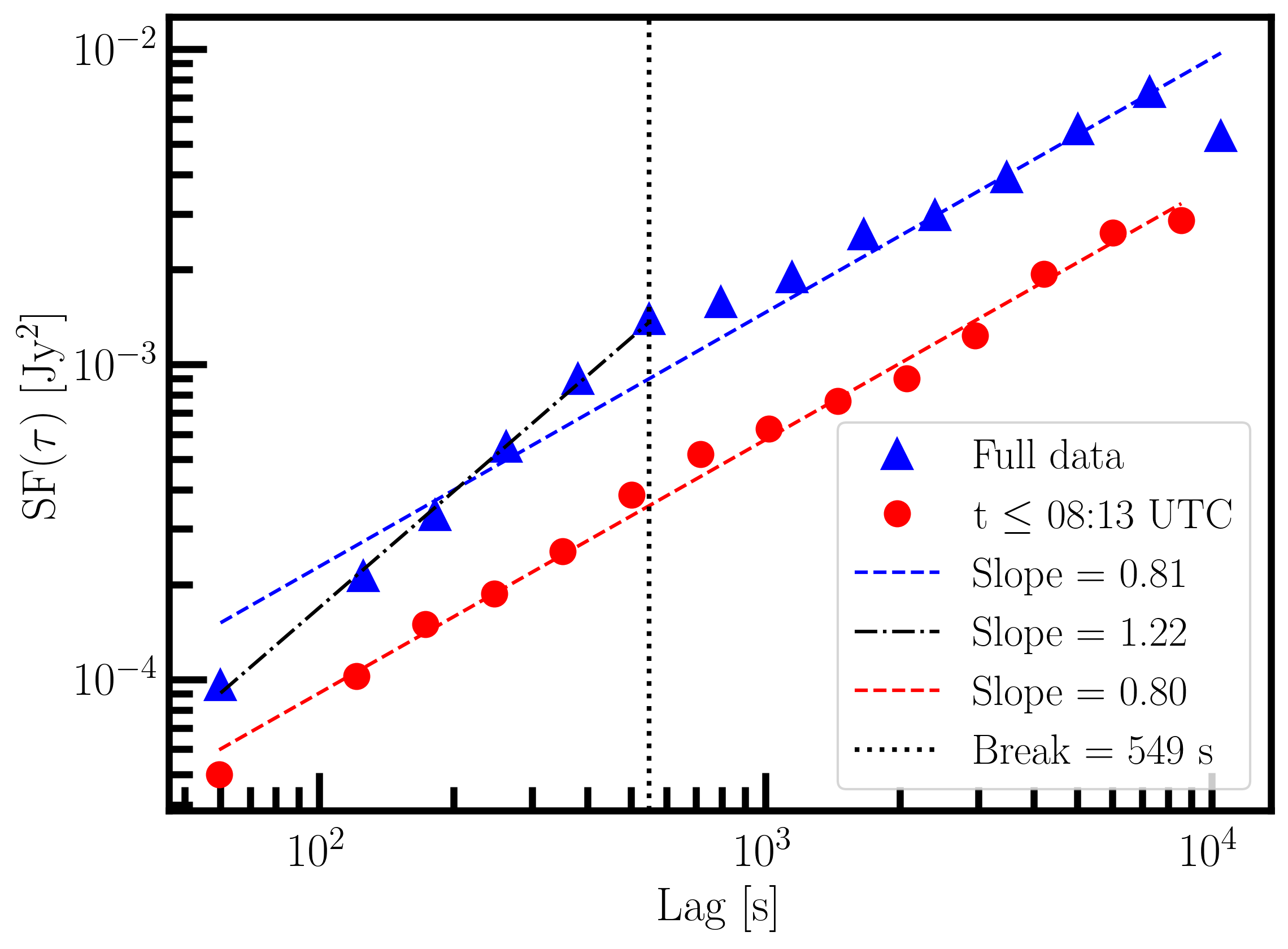} 
    \caption{SF of the full light curve (blue triangles) and t $\leq$ 08:13 UTC (red circles). The blue dashed line is the fit to the SF of the full light curve (slope = 0.81). The black dashed line is the fit to the first six data points (slope = 1.22) and the black dotted line indicates the tentative break at 549 s. The red dashed line is the fit to the SF of t $\leq$ 08:13 UTC light curve (slope = 0.80).}
    \label{fig:flux_density_sf}
\end{figure}

We determine the characteristic timescale of variability from the light curve using second-order structure function (SF) \citep[see][and references therein]{Iwata}. SF was calculated as ${\rm SF}(\tau) = \langle \left[S_\nu(t) - S_\nu(t + \tau)\right]^2\rangle$, where $\tau$ is the time-lag. Fig. \ref{fig:flux_density_sf} shows that the SF increases to $\sim$7259.75\,s ($\sim$2.02\,h) with a slope of $0.81 \pm 0.05$ (full; blue triangles) and $\sim$8556.28\,s ($\sim$2.38\,h) with a slope of $0.80 \pm 0.02$ (excluded; red circles), as expected for a red-noise process. The flux density distribution (Fig.~\ref{fig:flux_distribution_histogram}) (full) is asymmetric (Fisher--Pearson skewness of $-1.54$), with a peak around 0.84\,Jy and a tail extending towards lower flux densities (0.65--0.75\,Jy). In the SF of the full light curve (blue triangles in Fig.~\ref{fig:flux_density_sf}), we visually identify a tentative break at a timescale of $\approx 550$\,s, below which the slope is $1.22 \pm 0.03$. Although this timescale is broadly consistent with variability observed at millimetre wavelengths \citep{2022ApJ...930L..19W}, we argue that it is not representative of a persistent characteristic timescale. This interpretation is motivated by the presence of a distinct low-flux tail in Fig. \ref{fig:flux_distribution_histogram}, which becomes prominent beginning around 08:13\,UTC (highlighted by the pink shaded region in Fig.~\ref{fig:light_curve}; top panel). When this is excluded, the SF no longer shows a discernible break (red circles in Fig. \ref{fig:flux_density_sf}), and the flux density distribution of the remaining data is well described by a Gaussian (Fisher--Pearson coefficient of 0.45). This suggests that the tail in Fig. \ref{fig:flux_distribution_histogram} likely reflects a separate physical process, distinct from the main variations captured by the SF, and does not affect the determination of characteristic timescales.

If the accretion disk around Sgr\,A* is oriented close to a face-on view \citep{2022ApJ...930L..16E, 2022A&A...665L...6W}, the optical depth toward the innermost region may be significantly lower than what a 1-D model predicts and hence the 3\,GHz photosphere may be located closer to the black hole. In such a scenario millimetre and centimetre variability could have similar statistical characteristics, such that, the observed variability at low frequencies, in part, is caused by reprocessed emission from mechanism that drive variability at high frequencies. To test this, we compared the centimetre and millimetre variability by generating 50\,000 simulated light curves by drawing realizations from a Mat\'ern covariance Gaussian process whose parameters were fitted to the 230 GHz ALMA observations obtained in 2017 \citep{2022ApJ...930L..19W}. Each realization was sampled at a cadence of 60 s over a duration of 10 h and initially contained no measurement noise. For each synthetic light curve, we computed the SF after interpolating it onto the observed time grid using logarithmically spaced time-lag bins. To mimic realistic observational conditions, Gaussian noise with a standard deviation of 24\,mJy, estimated from the phase calibrator PKS 1830-3602, 
was added to these synthetic light curves. The slope of the SFs becomes shallower as a result of the addition of noise relative to the noise-free case (see Fig.~\ref{fig:flux_density_sf_hist}). The SF slope measured for the first MeerKAT light curve, $0.81 \pm 0.05$ (full), lies within the distribution obtained from the noise-added simulations, demonstrating  
consistency with the variability expected from resampling the ALMA light curves to MeerKAT’s cadence.  

\section{Discussion and Conclusion}
\label{sec:discussion}

This study represents the first use of MeerKAT to characterise Sgr\,A*'s variability and spectral properties. Compared to previous works that used widely spaced observations in time \citep[e.g.,][]{falcke1999}, our high-cadence monitoring allows new insights into short-timescale variability at low radio frequencies. Previous studies report the modulation index of 6\% at 8.3~GHz and 2.5\% at 2.3~GHz \citep{falcke1999} on timescale of weeks, whereas we measure 6.11\% on minute timescale (full) and 3.5\% (excluded), indicating that similar modulation is already present at shorter timescales.

Our dataset also shows notable peaks and troughs at intervals of $\sim$20 min (Fig.~\ref{fig:light_curve}; top panel), prominent from 06:25\,UTC onwards. Such short timescale variability is usually observed at higher frequencies \citep{genzel2003, baganoff,2022ApJ...930L..19W}. This suggests that the compact, event-horizon–scale source, known to dominate the millimetre emission of Sgr\,A* \citep{2022ApJ...930L..12E}, may also contribute to the variable component at radio frequencies near 3\,GHz. Another feature observed is the presence of sudden jumps in flux density, similar to those reported at higher NIR frequencies \citep{genzel2003}.  
As seen in Fig.\,\ref{fig:light_curve}, 
after 08:13\,UTC, the light curve exhibits significant changes in flux density. The origin is unclear, but may be due to absorption from the opacity effects in the Sgr~A* accretion flow. Alternatively, it could arise from out-flowing absorption material in the foreground, particularly if the system is observed at low inclination \citep{2022ApJ...930L..16E,2022A&A...665L...6W}. When considering the light curve up to 08:13 UTC, the bi-modality in the spectral index as a function of flux density is no longer present, indicating that the emission is predominantly optically thick.

One possible origin of variability seen for Sgr\,A* is from propagation effects of electromagnetic waves in the turbulent interstellar medium, e.g., via refractive and diffractive scintillation in the strong scattering regime. Following equations 10 and 11 from \cite{Psaltis:2018idv}, for Sgr\,A*, diffractive timescale ($t_{\rm dif}$) lies in the range 1 to 4\,s, and refractive timescale ($t_{\rm ref}$) in the range 17 to 86\,yr. These timescales at 10.74\,cm, the wavelength of our observations, were estimated assuming that the transverse velocities between the source and the scattering screen between 10 and $50~\mathrm{km\,s^{-1}}$, and the distance of potential scattering screen and observer of 2.7\,kpc. Since $t_{\rm dif}$ is smaller than our bin size of 1\,min, any diffractive scattering is effectively averaged out. $t_{\rm ref}$ is on the order of years, making it too slow to produce variability on smaller timescales. 

Observational evidence for interstellar scattering in the vicinity of Sgr~A* is provided by the radio magnetar J1745$-$2900, located just 3$\arcsec$ from Sgr\,A*, which exhibits interstellar scattering with a timescale of $\tau_{\rm sc} = 1.3 \pm 0.2$\,s at 1\,GHz \citep{2014ApJ...780L...3S}. As $\tau_{\rm sc} \propto \nu^{-4}$, this corresponds to $\tau_{\rm sc} < 1$\,s at 2.79\,GHz. Conservatively assuming that the scattering properties along the line of sight to Sgr~A* are comparable to those measured toward the nearby radio magnetar J1745$-$2900 our minute-scale binning is sufficient to wash out the diffractive effects. Hence, the observed variability of Sgr\,A* is unlikely to be caused by interstellar scattering.

Having ruled out interstellar scattering, we adopt a single zone model -- where a single physical process drives the variability across frequency bands, and perform the time lag analysis (Section~\ref{sec:time_lag}). We find a time delay of $826 \pm 97$\,s, with the light curve at 2.70\,GHz lagging behind that at 2.87\,GHz. This time lag is somewhat inconsistent to that predicted by the adiabatic expansion scenario in which the expanding blob of synchrotron emitting plasma near the vicinity of the accretion disk of Sgr~A* becomes optically thin, producing peaks in flux density that occur later at lower frequencies, naturally explaining the observed time delays \citep{2021MNRAS.505.3616M}. A similar approach was adopted by \cite{2022ApJ...930L..19W} at high frequencies (212--230 GHz), where adiabatic expansion models predict lags of order 1--2\,min. However, no significant delays were detected, with upper limits $\le 10$\,s. Taken together, these results highlight a frequency-dependent behaviour of Sgr\,A*, in which time-lag appears to differ between the centimetre ($\sim$3 GHz) and millimetre ($\sim$230 GHz) wavelengths. Although the SF slopes agree, the time lags are significantly different, which makes it a difficulty to present a consistent, comprehensive scenario. More data, over longer timescale to study the variability of Sgr~A* is needed.

The time delay between the two frequencies could also be determined in the context of the presence of a fast and directed outflow from a relativistic jet of Sgr~A* in which the peak frequency of radio emission changes with position along the jet axis \citep{refId0}. Based on the analysis by \citet{2015A&A...576A..41B}, the wavelength dependence of the time delay corresponds to a time-lag slope of $36\textrm{--}42\,{\rm min\,cm^{-1}}$ using 100~GHz ALMA data and 19--48\,GHz VLA data. For the frequency range between 2.70 and 2.87\,GHz, this time-lag slope predicts a lag of $\sim$$24\textrm{--}28$\,min. The measured delay of $\sim$$12\textrm{--}15$\,min is therefore inconsistent with the prediction based on a simplistic jet outflow model that directly relates Very Large Baseline Interferometry sizes and time lags.

The variation of spectral index contains useful information on changes in optical depth likely arising from turbulent accretion flow of plasma around Sgr~A* \citep{2022ApJ...930L..19W}. In Fig. \ref{fig:light_curve} (bottom panel), we report spectral index variation over time for Sgr~A*. Some fluctuations may also be due to systematic effects, such as the limited bandwidth, which can bias spectral index estimates. \citet{falcke1999} reported positive spectral index versus flux density correlation in the centimetre regime. Our spectral index remains independent for flux density $> 0.75$\,Jy, suggesting changes in flux density occur without affecting the spectrum, a behaviour also observed at NIR frequencies \citep{2007ApJ...667..900H, 2011A&A...528A.140T, 2014ApJ...796L...8W, 2024ApJ...977..228P}. If the radio emission is synchrotron, it is more plausible that the variability is driven by changes in magnetic field strengths and/or density of relativistic electrons. However, if the flux density change toward the end of the light curve arises from an absorption feature, this effect alone could explain the observed spectral-index variability. Future observations will help to test these interpretations.

From the SF analysis (Fig. \ref{fig:flux_density_sf}), we determine a characteristic timescale of $\sim$120 min. However, a precise determination of this timescale is difficult from a single light curve, since the SF can be affected by limited sampling. As described in \cite{falcke1999}, the characteristic timescales associated with Sgr\,A* at these frequencies, including those linked to black hole activity or relativistic outflows, are typically less than a day. Therefore, the timescale of $\sim$120 min could indicate global accretion flow instabilities. Additionally, \citet{farhad2011} reported flux variability on sub-minute to hour timescales in their radio data at 8, 22, and 43 GHz, which they attributed to the fact that most of the continuum flux from Sgr~A* at radio and sub-millimetre wavelengths is believed to originate within its accretion disk. The first light curve from MeerKAT reported in this letter indicates variability comparable to that at centimetre and millimetre wavelengths, where the measured SF slope is consistent with the expected values from ALMA light curves, whereas slower variability would have produced a much steeper slope. 

\begin{acknowledgements}
We acknowledge Farhad Yusef-Zadeh for initial discussions and helpful insights on this work. This work has made use of the MeerKAT radio telescope, which is operated by the South African Radio Astronomy Observatory, a facility of the National Research Foundation, an agency of the Department of Science and Innovation. This work has benefited from the ``MPIfR S-band receiver system”, designed, constructed, and maintained with funding from the MPI für Radioastronomie and the Max Planck Society. In particular, we acknowledge the contributions and efforts of the experts from the Electronics and Digital Signal Processing departments of the MPIfR. MW is supported by a Ramón y Cajal grant RYC2023-042988-I from the Spanish Ministry of Science and Innovation and acknowledges financial support from the Severo Ochoa grant CEX2021-001131-S funded by MCIN/AEI/ 10.13039/501100011033. SR acknowledges the support from the International Max Planck Research School for Astronomy and Astrophysics at the Universities of Bonn and Cologne. RSW and VVK acknowledge support from the European Research Council starting grant COMPACT (Grant agreement number: 101078094).
\end{acknowledgements}

\bibliographystyle{aa} 
\bibliography{bib} 

\begin{appendix}
\section{Observations and data reduction}
\label{sec:obs}
\subsection{Observing Strategy}
We used the MeerKAT radio telescope to observe Sgr\,A* using the S4 band receiver (2.62--3.50~GHz) on 21 March 2024 for an on-source time of 7.54\,h and a total bandwidth of 875\,MHz. These observations were performed as part of the ongoing MMGPS \citep{mmgps}. The data are observed in nine pointings dithering around Sgr\,A* (pointing centres shown in Table \ref{tab:pointings}), with the pointing GC00 is centred on the position of Sgr\,A*. The observation sequence began with a 5\,min scan of the primary calibrator, J$1939\textrm{--}6342$, followed by a 5\,min scan of the polarisation calibrator, 3C\,286, and 2\,min scans of the secondary (complex gain) calibrator, PKS\,$1830\textrm{--}3602$ interleaved between 20\,min scans on the different pointings of Sgr\,A*. Each of the pointings in Table~\ref{tab:pointings} were visited twice over the course of the observations run to ensure uniform $uv$-coverage. Another 5\,min scan of the primary calibrator was performed at the end of the observations.

\begin{table}[h]
\centering
\caption{Pointing positions used for the observations. Coordinates are given in J2000.}
\label{tab:pointings}
\begin{tabular}{@{}ccc@{}}
\toprule
\textbf{Pointing} & \textbf{RA (hh:mm:ss)} & \textbf{Dec (dd:mm:ss)} \\ \midrule
GC00              & 17:45:40.04            & -29:00:28.10            \\
GC01              & 17:46:02.51            & -29:00:28.10            \\
GC02              & 17:45:55.93            & -29:04:26.40            \\
GC03              & 17:45:40.04            & -29:06:05.10            \\
GC04              & 17:45:24.15            & -29:04:26.40            \\
GC05              & 17:45:17.57            & -29:00:28.10            \\
GC06              & 17:45:24.15            & -28:56:29.80            \\
GC07              & 17:45:40.04            & -28:54:51.10            \\
GC08              & 17:46:11.81            & -28:52:31.50            \\
\bottomrule
\end{tabular}
\end{table}

\subsection{Calibration and Band Selection}
\label{calibrationappendix}
We used the  MMGPS Imaging Pipeline to calibrate the data. We note that, the FWHM of MeerKAT's primary beam have complicated frequency-dependent jumps above 3\,GHz caused due to the excitation of TM11-mode within the waveguide of the orthomode transducers \citep{2023AJ....165...78D}. This also leads to jumps in the system equivalent flux density. These effects could affect flux density measurements of Sgr\,A*, particularly for pointings where it is not located at the phase centre. Hence, we discarded data above 3\,GHz in our analysis, limiting the bandwidth to 230\,MHz at a central frequency of 2.79\,GHz. To analyse spectral properties, we sub-divided the available bandwidth into four sub-bands of 57\,MHz with central frequencies at 2.70, 2.76, 2.81 and 2.87\,GHz.

\subsection{Point Source Fitting}

The self-calibrated data set was divided into 1\,min intervals using the \textsc{CASA} \citep{2022PASP..134k4501C} tool \texttt{mstransform} to examine the variability of the flux density of the source. We then employed the \texttt{uvmodelfit} tool \citep{marti} in \textsc{CASA} to perform a single-component model fitting to the \textit{uv}-data, estimating the flux density of the source.

\subsection{Source Isolation and Primary Beam Correction}

Given the complexity of the GC environment, characterized by diffuse emission and the presence of the mini-spiral around Sgr\,A*, we limit our analysis to the large $uv$-distances $\ge 5\times 10^4\,\lambda$ to isolate the emission from Sgr\,A*. This corresponds to sensitivity to structures smaller than $\sim$$4\arcsec$, to ensure that only compact emission, dominated by Sgr\,A*, was included. To further ensure that the \textit{uv} coverage are the same for all frequencies in our analysis, we used data in the \textit{uv}-range 50 000 to 67 403.3$\lambda$.

In order to correct for the primary beam (PB), we generated a PB model at the central frequency of 2.79\,GHz using \texttt{KATBEAM} \citep{katbeam}. Although the PB remains the same at a particular frequency, the relative position of the source within the beam shifts over time as a result of the dithering of the pointings. To account for this variation, we simulate the motion of the source along a circular trajectory centred on the phase centre at a fixed radial distance. This simulation enabled us to evaluate how minor positional shifts influence the flux correction changes based on the PB response. Specifically, we sampled 100 points along the circular path and estimated the PB correction factor as a function of the azimuth angle for each point. We then calculated the coefficient of variation (standard deviation divided by the mean) of the PB values and incorporated this into flux density uncertainty. This procedure was performed for all four sub-bands.

\section{Flux distribution, the light curves of the sub-bands and cross-correlation analysis} \label{appendix B}

\begin{figure}[h]
    \centering
    \includegraphics[width=\linewidth]{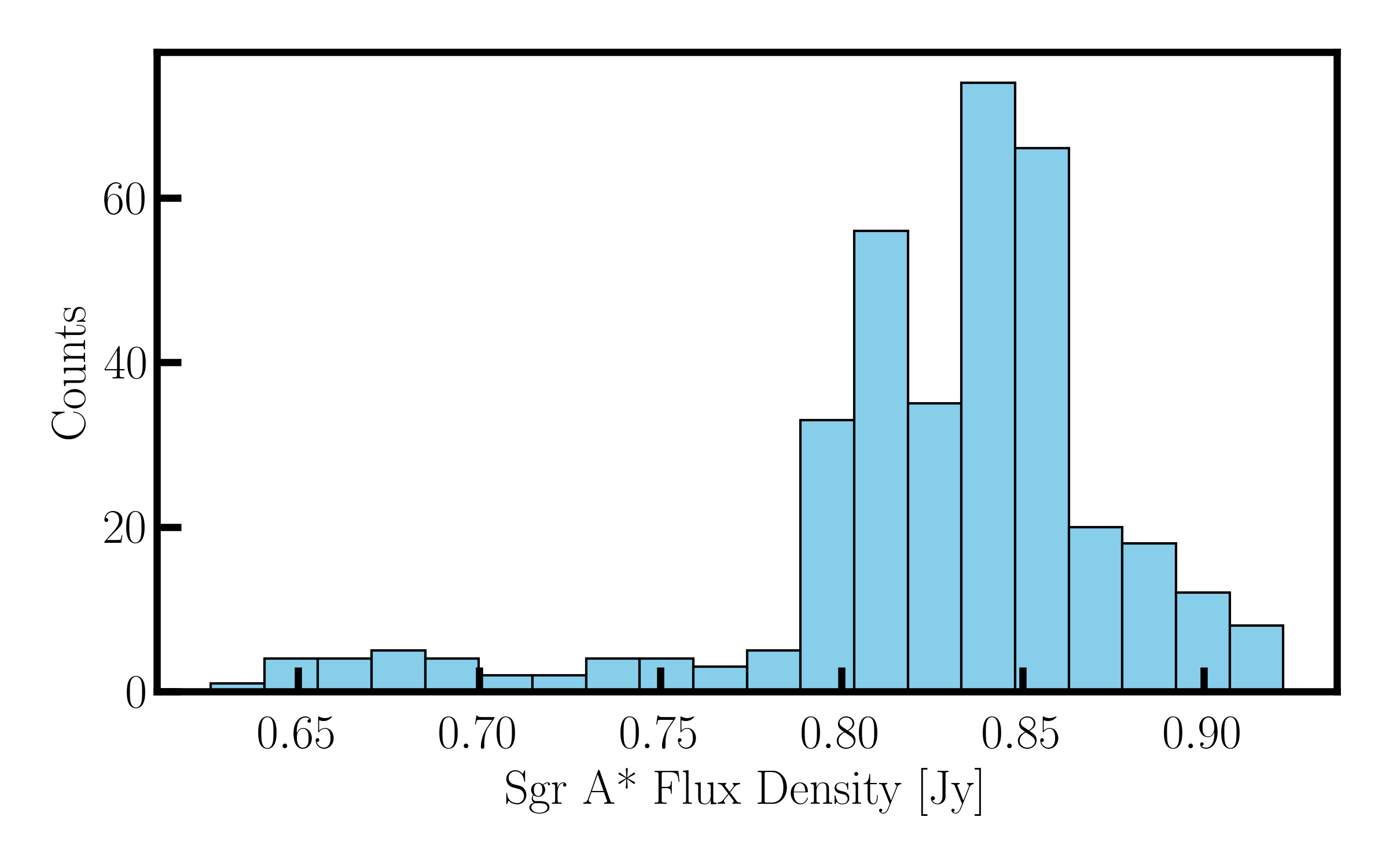} 
    \caption{Distribution of flux densities  of Sgr\,A*
    at 2.79\,GHz in bins of 1\,min. The distribution exhibits a distinct maximum at comparatively high flux densities, with an extended tail toward lower values spanning the range of observed flux densities.}
    \label{fig:flux_distribution_histogram}
\end{figure}

\begin{figure}[htbp]
    \centering
    \includegraphics[width=\linewidth]{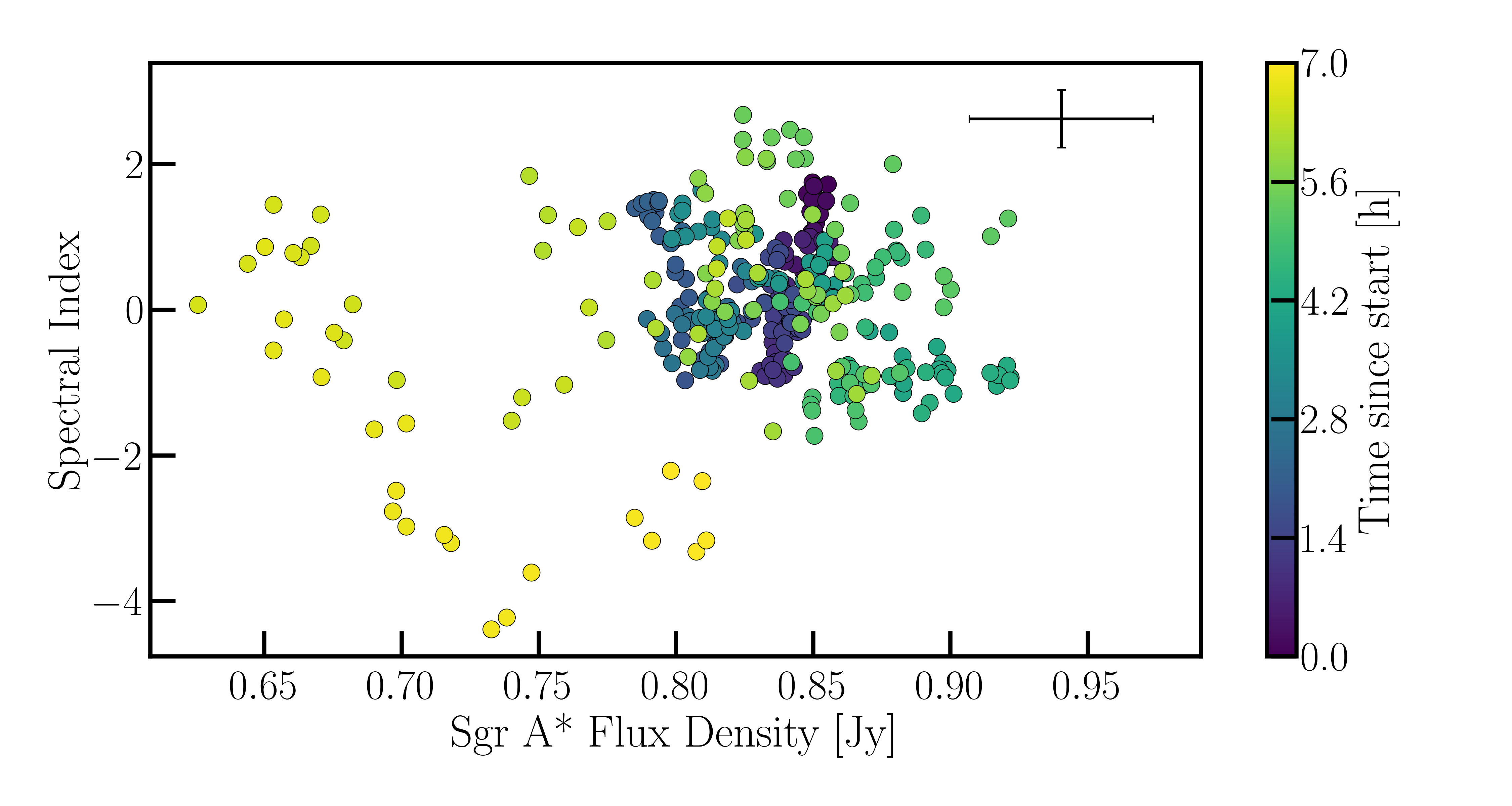} 
    \caption{Variation of spectral Index ($\alpha$) as a function of flux density of Sgr\,A* at 2.79\,GHz.}
    \label{fig:spectral_index_vs_flux}
\end{figure}

\begin{figure}[htbp]
   \centering
   \includegraphics[width=\linewidth]{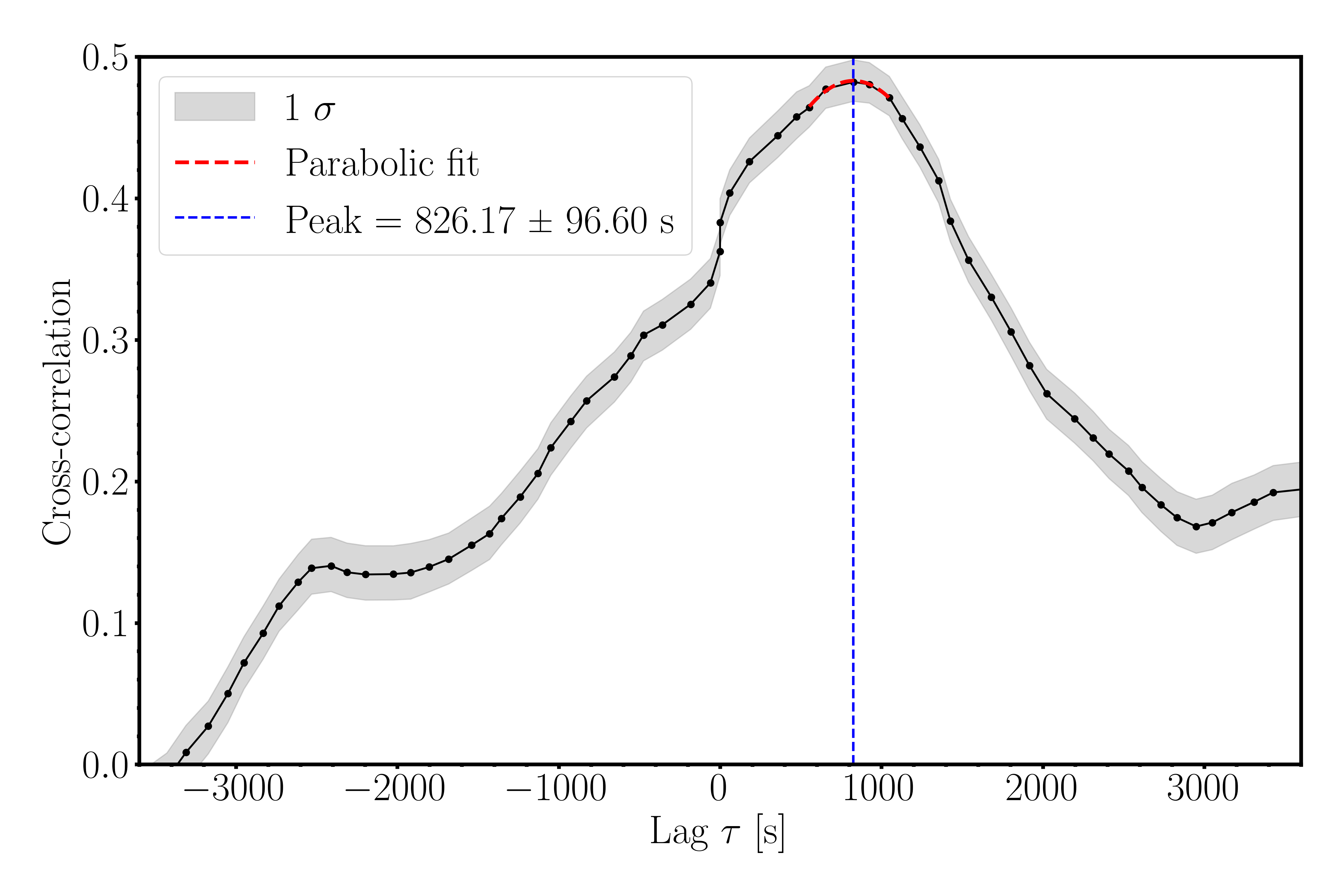} 
   \caption{The cross-correlation amplitude as a function of time lag ($\tau$) between light curves at 2.87 and 2.70\,GHz. The gray shaded band represents the 1\,$\sigma$ uncertainties on the box-car smoothed cross-correlation function.}
    \label{fig:time_lag_parabola}
\end{figure}

\begin{figure}[htbp]
    \centering
    \includegraphics[width=\linewidth]{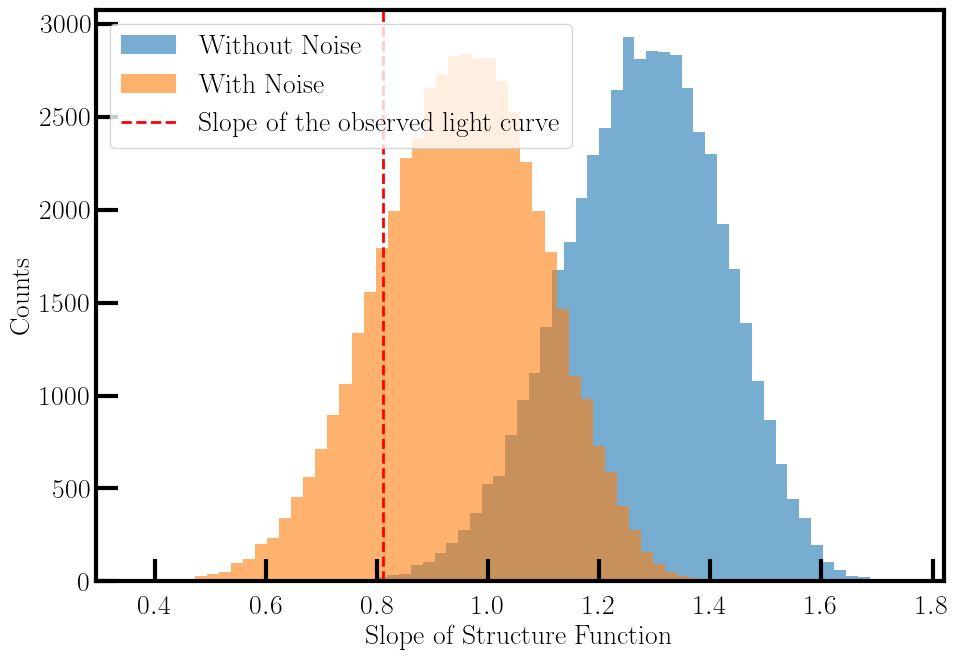} 
    \caption{Histogram of the slopes of the SF for 50 000 synthetic light curves, with and without added noise. The red line marks the slope derived from the observed light curve, which falls within the expected range.}
    \label{fig:flux_density_sf_hist}
\end{figure}

Here we present the light curves of Sgr\,A* at 2.70, 2.76, 2.81 and 2.87 GHz (Fig. \ref{fig:light_curve_sub-band}), the flux density distribution of the light curve at 2.79 GHz (Fig. \ref{fig:flux_distribution_histogram}), spectral index as a function of flux density (Fig. \ref{fig:spectral_index_vs_flux}), the cross-correlation plot of the highest (2.87 GHz) and the lowest (2.70 GHz) frequency sub-bands (Fig. \ref{fig:time_lag_parabola}), and the histogram of the SF slopes (Fig. \ref{fig:flux_density_sf_hist}). 

The cross-correlation\footnote{Cross-correlations were performed using the \texttt{pyZDCF} package \citep{Jankov2022}, a Python implementation of the Z-transformed Discrete Correlation Function \citep{1997ASSL..218..163A}.} was calculated as a function of time lag, using a binning scheme with a minimum of 11 points per bin and a maximum lag sampling at half the total duration of the light curve. To assess uncertainties, we performed 1000 Monte Carlo realizations of the cross-correlation curve by randomly perturbing its values within their measurement errors. Fig. \ref{fig:time_lag_parabola} shows the cross-correlation function between 2.87 GHz and 2.70 GHz light curves. We determine a time delay of $826 \pm 97$\,s by fitting a second-order polynomial to the peak, obtained by smoothing the cross-correlation function with a boxcar function corresponding to a width of 10~bins. The uncertainty corresponds to the standard deviation of the peak time lags derived from the 1000 Monte Carlo realizations of the smoothed cross-correlation function, generated by propagating the measurement errors provided by \texttt{pyZDCF}.

\begin{figure}[h]
    \centering
    \includegraphics[width=\linewidth]{lightcurve_subplot.png} 
    \caption{Total intensity light curves for Sgr A* for four sub-band frequencies (1 min time bins). The light curve of each sub-band is offset by +0.2 onward from the first sub-band in the flux density for visualization purposes. The error bars include systematic and $1\sigma$ statistical uncertainties. The light curves show similar overall trends, with the highest-frequency sub-band (2.87 GHz) leading the lowest-frequency sub-band (2.70 GHz).}
    \label{fig:light_curve_sub-band}
\end{figure}

\section{Verification tests for Sgr A* light curve} \label{appendix:tests}

\subsection{Test I: Light Curve of the Calibrator} \label{appendix:C1}

We generated the light curve of the secondary calibrator PKS\,1830$-$3602 using the same method as that for Sgr\,A* shown in Fig.~\ref{fig:light_curve} (top panel). Since the flux density of PKS\,1830$-$3602 remained relatively stable throughout the observation, any observed trends in the target’s light curve cannot be attributed to instrumental effects, thereby validating our approach. We determine the average flux density of the secondary calibrator to be 
$2957\pm118$\,mJy (where the mean statistical and mean systematic uncertainties are added in quadrature) at 2.79\,GHz. It is close to the flux density of PKS\,1830$-$3602 ($\sim$2830~mJy) between 2.877 and 2.962\,GHz \footnote{\label{skawiki}See: \url{https://skaafrica.atlassian.net/wiki/spaces/ESDKB/overview?homepageId=41025669}} with a measured modulation index of 0.8\%.

We also conducted a spectral index analysis for our secondary calibrator and found it to be relatively constant with a mean spectral index of -1.379 $\pm$ 0.004. The variation in the spectral index, as shown in Fig.~\ref{fig:spectral_index_combined}, 
is minimal ($-1.44$ to $-1.34$ in \textit{uv} domain) compared to the significant variation observed for our source, Sgr~A*, whose spectral index ranges between $-4$ to $+2$.

\subsection{Test II: Image domain vs. \textit{uv} domain}
We validated our visibility-domain methodology by conducting a similar analysis in the image domain for the secondary calibrator. We imaged all 1 min interval measurement set files using \textsc{WSClean} \citep{2014MNRAS.444..606O}, applying the same \textit{uv} range and dividing the data into four sub-bands to facilitate spectral index analysis. This approach proved more suitable for the secondary calibrator, which is bright, phase-centred, and situated in a less extreme environment than Sgr A*.

\begin{figure}[htbp]
    \centering
    \includegraphics[width=\linewidth]{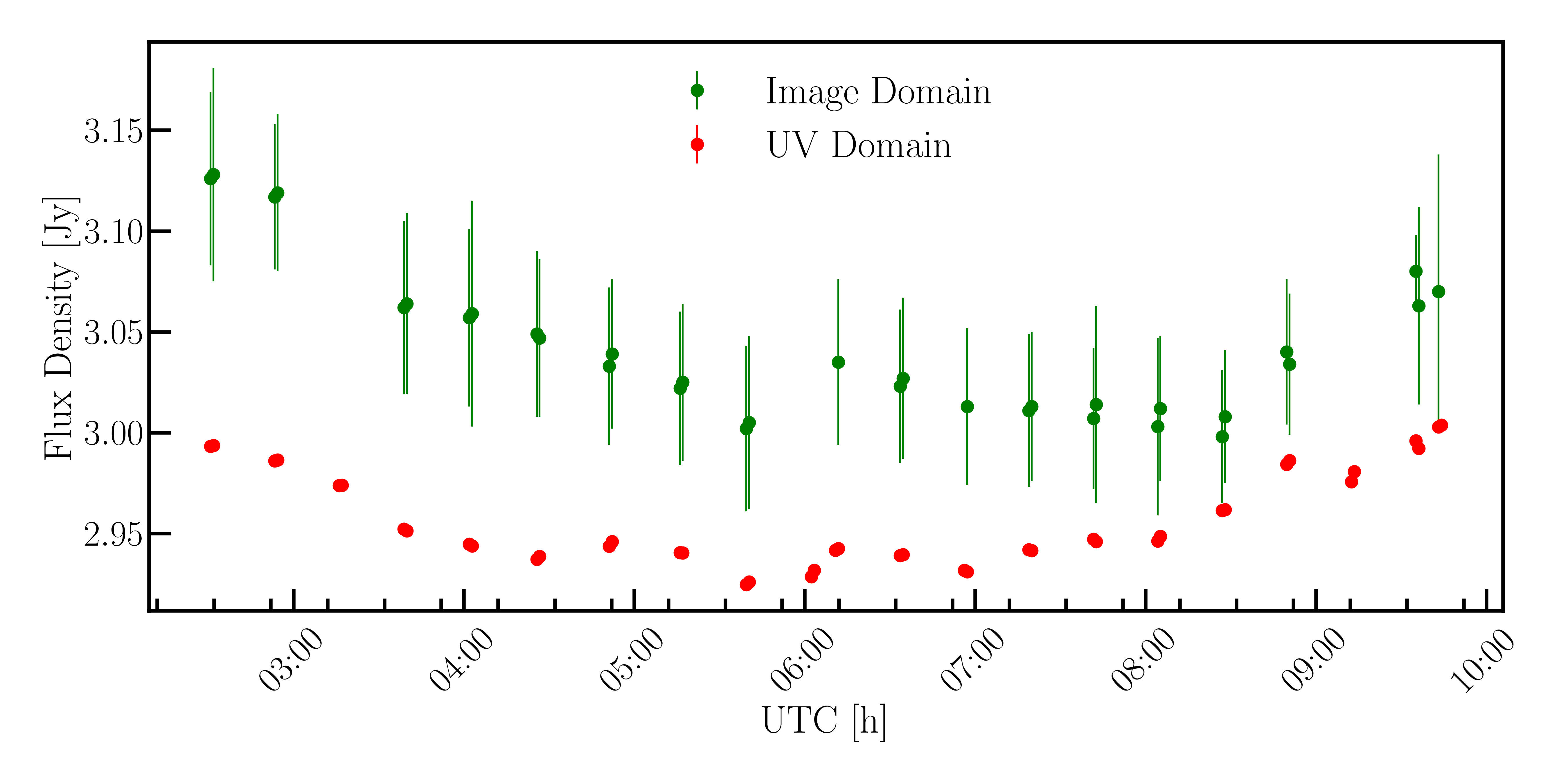} 
    \caption{Comparison of flux density light curves for the secondary calibrator PKS 1830–3602 in the image and visibility domains. We used \texttt{imfit} on images generated in 1 min bins to obtain the green curve, which represents the image-domain flux densities with associated errors. The red curve shows flux densities derived using \texttt{uvmodelfit} in the visibility domain. The error bars represent the $1\sigma$ statistical uncertainty. We observe consistent trends between the two methods, validating the robustness of our approach. Differences in error magnitudes reflect the distinct modelling assumptions in each domain. Image-domain errors are larger due to contributions from deconvolution and beam uncertainties, while visibility-domain errors remain smaller because they arise from direct model fitting in the Fourier domain.}
    \label{fig:flux_density_comparison}
\end{figure}

\begin{figure}[htbp]
    \centering
    \includegraphics[width=\linewidth]{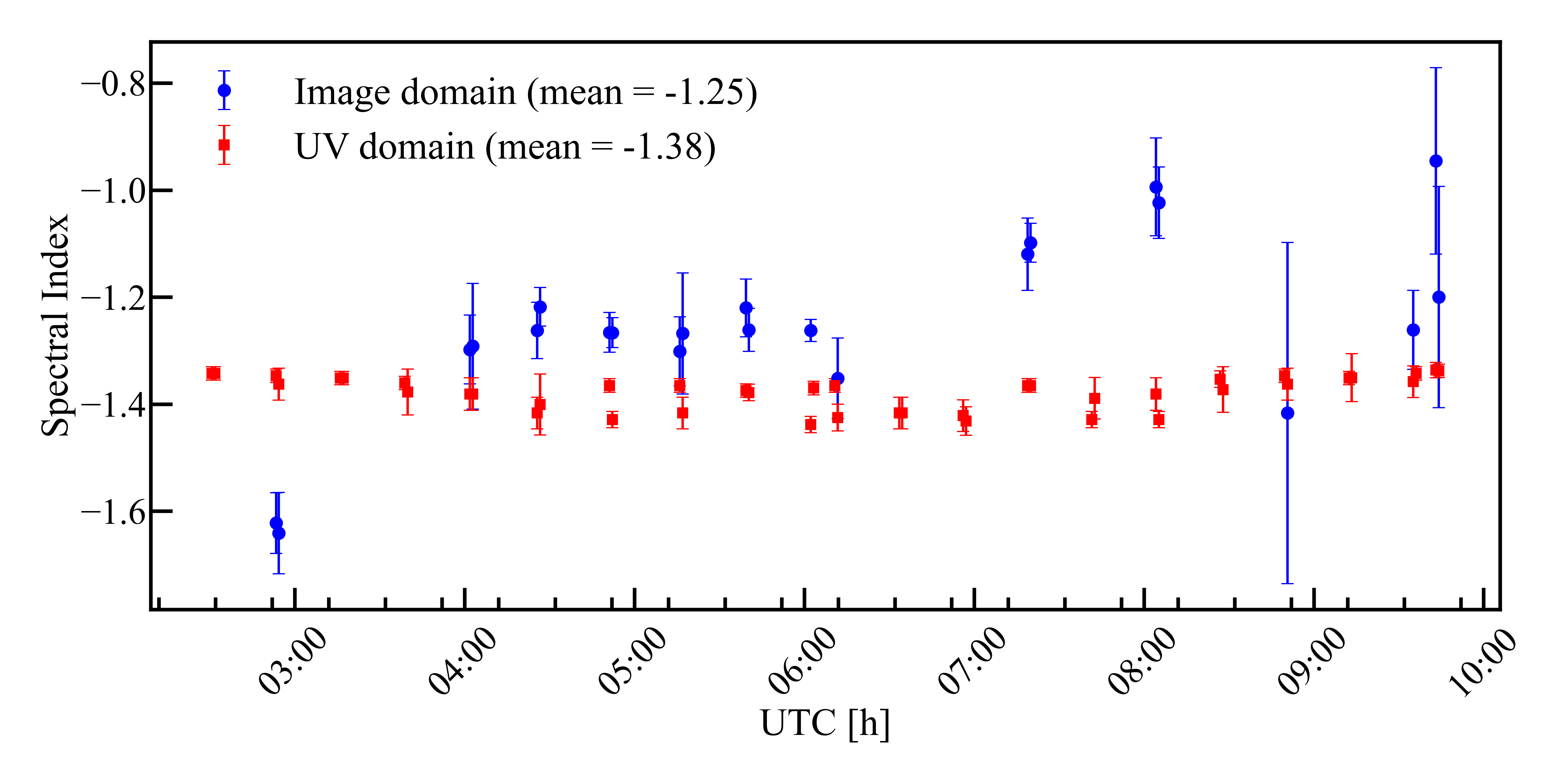} 
    \caption{Spectral index ($\alpha$) variation of PKS 1830-3602 over time, derived from sub-band flux measurements. Error bars reflect statistical uncertainties.}
    \label{fig:spectral_index_combined}
\end{figure}

After generating the images, we used the \textsc{CASA} task \texttt{imfit} to extract flux densities and corresponding errors. The resulting light curve appears in Fig. \ref{fig:flux_density_comparison}, and the spectral index variation over time is shown in Fig. \ref{fig:spectral_index_combined}. Light curves from both the visibility and image domains exhibit consistent trends. The mean flux density is 3042 $\pm$ 129~mJy (where the mean statistical and mean systematic uncertainties are added in quadrature), and the mean spectral index is -1.25 $\pm$ 0.02.

This cross-domain comparison also enables a more accurate assessment of error values. Errors in the image domain are $\sim$100 times smaller than the measured flux density, whereas, in the visibility domain, they are about 10 000 times smaller. The discrepancy likely stems from fewer degrees of freedom in the image domain, where multiple sources are modelled, compared to the single-source model used by \texttt{uvmodelfit} in the visibility domain. Furthermore, image-domain errors include contributions from deconvolution and cleaning, which are absent in the visibility domain.

Discrepancies in flux density values may arise because \texttt{imfit} sometimes fails to properly deconvolve the source from the beam, classifying it as a point source without a reliable size estimate in many snapshots. Gaps in the light curve correspond to failed \texttt{imfit} attempts, often caused by extreme beam elongation that leads to World Coordinate System (WCS) conversion errors, likely due to numerical instability. In particular, the snapshot images for timestamps 9:12:25.5 and 9:13:25.5 are missing, indicating that \textsc{WSClean} did not generate them, and further investigation is needed. The reduced number of data points in the spectral index plot compared to the visibility domain results from \texttt{imfit} failures in individual sub-bands, which prevented spectral index calculation at those times.

\subsection{Test III: Light curves of other sources in the field}
\begin{figure}[htbp]
    \centering
    \includegraphics[width=\linewidth]{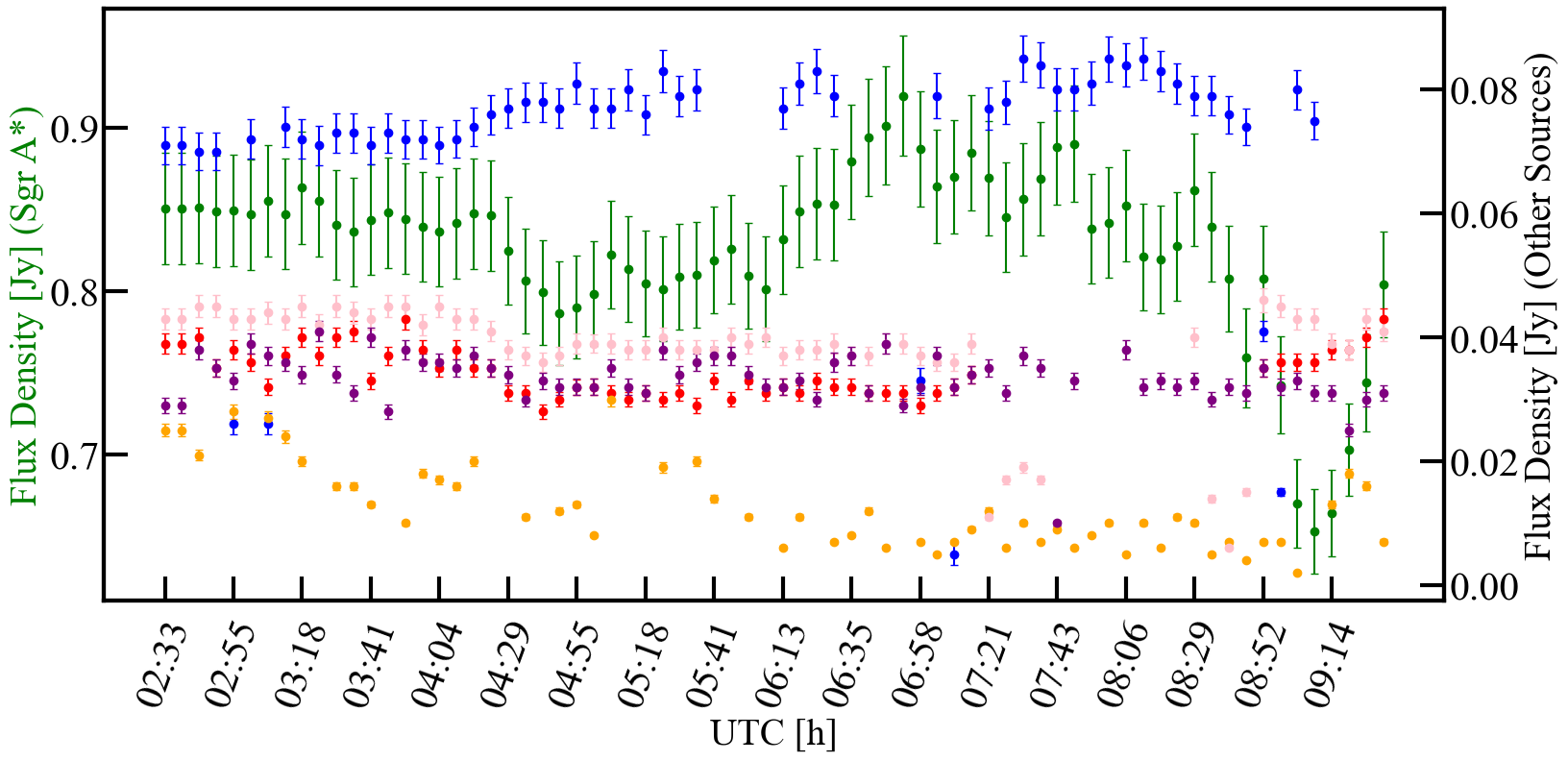} 
    \caption{Total intensity light curve of Sgr\,A* in bins of 5\,min (green dots) and the the other point sources
    ({blue}: J1744-2857, {red}: J1746-2845, {orange}: J1745-2904, {pink}: J1746-2856, {purple}: J1746-2852) on 21 March 2024 at 2.79~GHz. The error bars include systematic and $1\sigma$ statistical uncertainties.}
    \label{fig:lightcurve_radio_sources}
\end{figure}

The analysis in Appendix \ref{appendix:C1} demonstrates that the phase calibration solutions have not caused any instrumental time variability. However, it does not rule out any other potential issue restricted only to the target scans. The phase calibrator is always observed at the pointing centre, but Sgr A* is offset from it in most of the GC pointings. The target scans are also self-calibrated, while the phase calibrator is not. Hence, we repeated the analysis for some off-centre point sources in the field of view. For this, we made light curves of these sources, but this time with 5 min binning and visibilities $>$ 30 000$\lambda$. This was done in order to gain sufficient signal-to-noise to apply the same \texttt{uvmodelfit} technique as these sources are $\sim$10 times weaker than Sgr~A*. The light curves of these sources, along with Sgr~A*, with 5 min binning are shown in Fig. \ref{fig:lightcurve_radio_sources}.

The comparison of 5 min Sgr~A* light curve with those of other point sources in the GC shows that none of the point source follows the shape of the light curve obtained for Sgr~A*. There are some drop-off points and gaps in the time series for some GC pointings of these sources, indicating that the \textit{uv}-modelling failed, possibly because the sources were right at the edge of the PB in these cases.

From this analysis, the important conclusions are that the short-timescale variations are not seen in other point sources in the GC fields, and the level of fluctuations is smaller than that of Sgr~A* even though PB-correction related errors in these sources are relatively higher. Also, based on the phase-calibrator light curve (Fig. \ref{fig:light_curve}; top panel), the short-timescale fluctuations due to incorrect amplitude solutions can be ruled out. We also do not see jumps in flux densities of Sgr~A* or the point sources across scans. Lastly, the flux densities estimated for 5 min binning for Sgr~A* follow the 1 min time bins right on top of each other. All the tests presented in this Appendix support the conclusion that  the observed light curve variability is unlikely to be caused by instrumental effects related to the dithering of Sgr A* across target scans.

%
%
%

\end{appendix}

\end{document}